# Some properties of the central heavy ion collisions

*Z. Wazir*[*,1], *M. K. Suleymanov*[1,2], *E. U. Khan*[1], *Mahnaz Q. Haseeb*[1], *M. Ajaz*[1], *K. H. Khan*[1]

[1]CIIT, Islamabad (Pakistan), [2]JINR, Dubna (Russia)

*Corresponding author: zafar_wazir@comsats.edu.pk

## Abstract

Some experimental results are discussed in connection with the properties of the central heavy ion collisions. These experiments indicate the regime changes and saturation at some values of the centrality. This phenomenon is considered to be a signal of the percolation cluster formation in heavy ion collisions at high energies.

## 1. Introduction

Study of the centrality dependence of the characteristics of hadron-nuclear and nuclear-nuclear interactions is an important experimental way for obtaining information on phases of strongly interacting matter formed during the collision evolution. L. Wan Hove was first in attempting to use the centrality to get information on the new phases of matter [1] using the data coming from the ISR CERN experiments on pp-interactions. To fix the centrality the particle density ($\Delta n$) in given region of rapidity was considered ($\Delta y$) -- $\Delta n/\Delta y$. The ISR data shows that by increasing the values of the $\Delta n/\Delta y$ starting from some values of the $\Delta n/\Delta y$ the $p_t$ distribution becomes wider (see Fig.1). Wan Hove aimed to explain the fact as a signal on deconfinement [2] in hot medium and formation of the Quark Gluon Plasma (QGP) [3]. In this paper we discuss some properties of central collisions. These are necessary to get the signal on the deconfinement and to identify QGP.

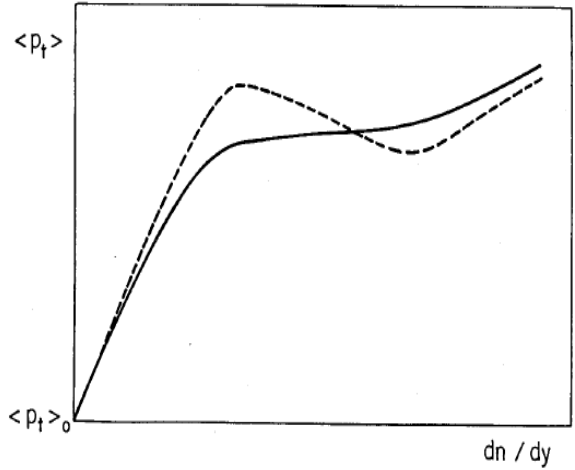

**Fig. 1.** $P_t$ vs. n Correlation could provide a signal for the deconfinement transition of hadronic matter.

# 2. Some properties of central collisions

## 2.1 Hadron-Nucleus Collisions

Fig.2 demonstrates a number of $\pi^{12}C$-interactions ($N_{star}$) as a function of the number of identified protons $N_p$ [4]. In this experiment $N_p$ was used to fix the centrality. One can see the regime change in the behavior of the values of $N_{star}$ as a function of $N_p$ near the value of $N_p=4$. The value was used to select the $\pi^{12}C$-reactions with total disintegration of nuclei (central collisions).

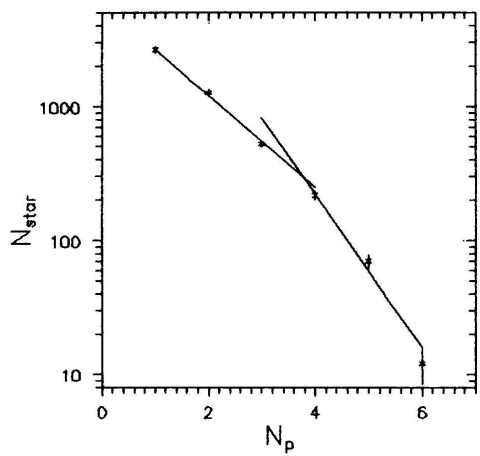

**Fig.2** A number of $\pi^{12}C$-interactions (at $P_{\pi^-}$=40 GeV/c) as a function of the number of identified protons.

As an example of existence of the regime change in proton-nucleus collisions we can show $\Lambda$ production (see Fig.3) as a function of collision centrality for 17.5 GeV/c $p$–$Au$ collisions has been measured by BNL E910 [5]. The centrality of the collisions is characterized using a derived quantity $\nu$, the number of inelastic nucleon-nucleon scatterings suffered by the projectile during the collision. The open symbols are the integrated gamma function yields, and the errors shown represent 90% confidence limits including systematic effects from the extrapolations. The full symbols are the fiducial yields. The various curves represent different functional scaling. The same results have been obtained by BNL E910 Collaboration for $\pi^-$-, $K^0_s$- and $K^+$- mesons emitted in $p+Au$ reaction.

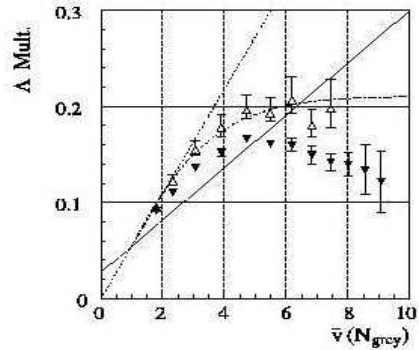

**Fig.3** The $\Lambda$ yield versus $v$.

## 2.2 Nucleus-Nucleus Collisions

From Fig.4 one can see the behaviors of the event number as a function of centrality for light nuclei interactions: *dC*-, *HeC*- and *CC*-interaction at 4.2 A GeV/c [6]. There are regime changes again for these interactions. These points of regime change could be used to select the central collisions.

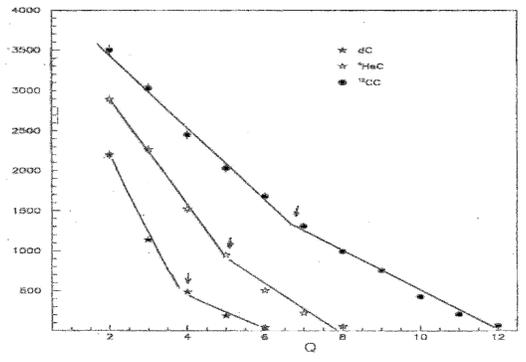

**Fig.4** The number of *dC-, HeC-* and *CC-* reactions as a function of centrality at 4.2 A GeV/c.

## 2.3 Heavy Ion Collisions

Experimental ratios of $<K^+>$, $<K^->$, $\varphi$, and $\Lambda$ to $<\pi^{\pm}>$ plotted as a function of system size (Fig.5). Statistical errors are shown as error bars, systematic errors if available as rectangular boxes. The curves are shown to guide the eye and represent a functional form

$$a - b \exp( -<N_{part}> /40).$$

At $<N_{part}>=60$ they rise to about 80% of the difference of the ratios between $N_{part} =2$ and 400 [7].

The ratio of the $J/\psi$ to Drell-Yan cross-sections has been measured by NA38 and NA50 SPS CERN (see Fig.6) as a function of centrality of the reaction estimated, for each event, from the measured neutral transverse energy $E_t$ [8]. Whereas peripheral events exhibit the normal behavior already measured for lighter projectiles or targets, the $J/\psi$ shows a significant anomalous drop of about 20 % in the $E_t$ range between 40 and 50 GeV. A detailed pattern of the anomaly can be seen in Fig. 6 which shows the ratio of the $J/\psi$ to the Drell-Yan cross-sections divided by the exponentially decreasing function accounting for normal nuclear absorption. Other significant effect which is seen from this picture is a regime change in the $E_t$ range between 40 and 50 GeV both for light and heavy ion collisions and saturation.

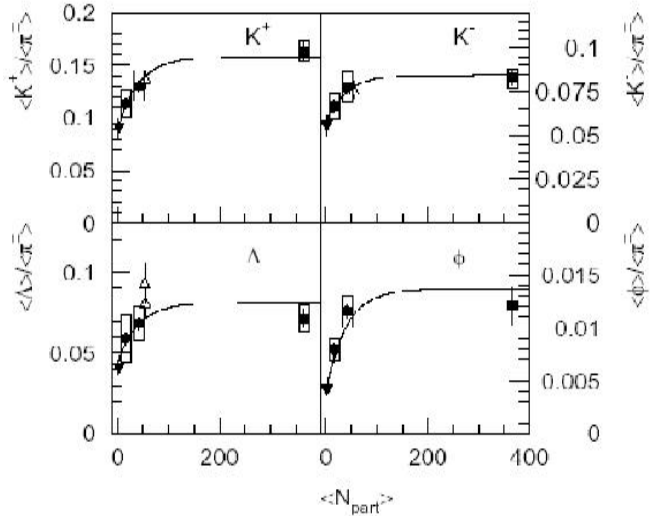

**Fig.5** The experimental ratios of $<K^+>, <K^->, \varphi$, and $\Lambda$ to $<\pi^{\pm}>$ plotted as a function of system size (▼ $p+p$, $C+C$ and $Si+Si$, ● $S+S$, ■ $Pb+Pb$).

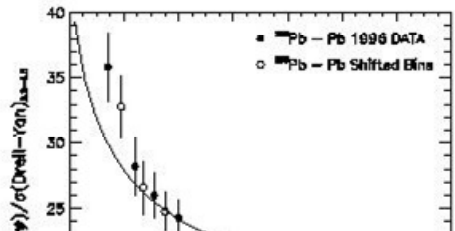

**Fig.6.** The ratio of the *J/ψ* to Drell-Yan cross-sections as a function of centrality.

Recent data obtained by STAR RHIC BNL[9] on the behavior of the nuclear modification factors of the strange particles as a function of the centrality in *Au+Au*- and *p+p*-collisions at $\sqrt{s_{NN}}$ = 200 GeV is shown in Fig.7. One can see regime change and saturation for the behavior of the distributions. To fix the centrality the values of participants ($N_{part}$) was used.

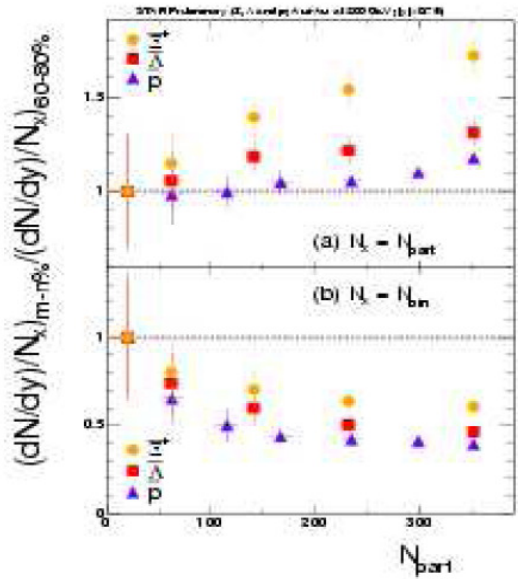

**Fig.7.** The nuclear modification factors of the strange particles as a function of centrality.

Recent results from RHIC on heavy flavor production [10] show nuclear modification function

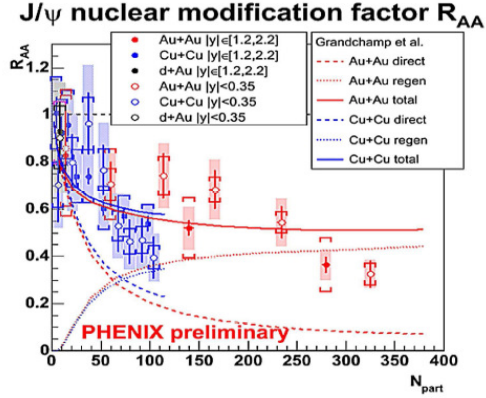

**Fig.8** The values $R_{AA}$ as a function of $N_{part}$ for heavy flavor coming from RHIC.

($R_{AA}$) distributions for *Au+Au* and *Cu+Cu* collisions (see Fig.8) as a function of centrality. A number of participants ($N_{part}$) were used to fix the centrality. We can see again the regime change and saturation in the behavior of the distributions.

## 3. Discussion

The points of regime change appear for the behavior of some characteristics of events as a function of centrality as some critical phenomena for hadron-nuclear, nuclear-nuclear interactions and for ultrarelativistic ion collisions. The phenomenon observed in the wide range of energy and almost for all particles (mesons, baryons, strange particles and heavy flavor particles). After point of regime change the saturation is observed. The simple models (such as wounded-nucleon model and the cascade model) which are usually used to describe the high energy hadron-nuclear and nuclear-nuclear interactions could not explain the results. To explain this result it is necessary to suggest that the dynamics of the phenomena is same for hadron-nuclear, nuclear-nuclear and heavy ion interactions and is independent of the energy and mass of the colliding nuclei. The responsible mechanism to describe the above mentioned phenomena could be statistical or percolation ones because phenomena have a critical character. In Ref. [11] complete information was presented about using statistical and percolation models to explain the experimental results coming from heavy ion physics. However, it is known that the statistical models give more strong A-dependences than percolation mechanisms. That is why we believe that the responsible mechanism to explain the phenomena could be percolation cluster formation [12]. Big percolation cluster may be formed in the hadron-nuclear, nuclear-nuclear and heavy ion interactions independent of the colliding energy. But the structure and the maximum values of the reached density and temperature of hadronic matter could be different for different interactions depending on the colliding energy and masses within the cluster. Ref. [13] shows that deconfinement is expected when the density of quarks and gluons become so high that it no longer makes sense to partition them into color-neutral hadrons, since these would strongly overlap. Instead we have clusters much larger than hadrons, within which color is not confined; deconfinement is thus related to cluster formation. This is the central topic of percolation theory, and hence a connection between percolation and deconfinement [13] seems very likely. So we can see that the deconfinement could occur in the percolation cluster. Ref. [13] explains the charmonium suppression as a result of deconfinment in cluster.

Experimental observation of the effects connected with formation and decay of the percolation clusters in heavy ion collisions at ultrarelativistic energies and the study of correlation between these effects could provide the information about deconfinement of strongly interacting matter in clusters. We suggest two effects to identify the percolation cluster formation. One is the nuclear transparency effect and other light nuclear production.

In Ref. [14] percolation cluster is a multibaryon system. Increasing the centrality of collisions, its size and masses could increase as well as its absorption capability and we may see saturation. So after point of regime change the conductivity of the matter increases and it becomes a superconductor [13] due to the formation of percolation clusters. In such systems the quarks must be bound as a result of percolation.

The critical change of transparency could influence the characteristics of secondary particles and may lead to their change. As collision energy increases, baryons retain more and more of the longitudinal momentum of the initial colliding nuclei, characterized by a flattening of the invariant particle yields over a symmetric range of rapidities, about the center of mass - an indicator of the onset of nuclear transparency. To confirm the deconfinement in cluster it is necessary to study the centrality dependence in the behavior of secondary particles yields and simultaneously, critical increase in transparency of the strongly interacting matter.

Appearance of the critical transparency could change the absorption capability of the medium and we may observe a change in the heavy flavor suppression depending on their kinematical characteristics. It means that we have to observe the anomalous distribution of some kinematical parameters because those particles which are from area with superconductive properties (from cluster) will be suppressed less than the ones from noncluster area. So the study of centrality dependence of heavy flavor particle production with fixed kinematical characteristics could provide the information on changing of absorption properties of medium depending on the kinematical characteristics of heavy flavor particles.

In. Ref.[15] it was suggested that the investigation of the light nuclei production as a function of the centrality could give the clue on freeze-out state of QGP formation, which may be used as an additional information to confirm the percolation cluster formation near the critical point. There are two types of light nuclei emitted in heavy ion collisions: first type are the light nuclei which get produced as a result of nucleus disintegration of the colliding nuclei; while the second ones are light nuclei which are get comprised of protons and neutrons (for example, as a result of coalescence mechanism) which were produced in heavy ion interactions. In an experiment we can separate these two types of nuclei from each other using the following ideas: the yields for first type of nuclei have to decrease, by some regularity, with centrality of collisions. On the other hand, formation of the clusters could be a reason of the regime change in the behaviour of light nuclei yields as a function of centrality in the second type.